\pdfoutput=1
\documentclass[a4paper]{article}
\usepackage{ISCSLP2026}
\usepackage{ifthen}

\newboolean{blind}
\setboolean{blind}{false}

\usepackage{mathptmx}       % Times Roman for text + math (CFP requirement; pairs with ISCSLP sty)
\usepackage[scaled=0.92]{helvet}  % Sans for headings (Helvetica family)
\usepackage{courier}        % Mono
\usepackage{graphicx}
\usepackage{booktabs}
\usepackage{amsmath, amssymb}
\usepackage{xcolor}
\usepackage[hidelinks]{hyperref}
\hypersetup{
  pdftitle={Causal Localization of the Refusal Direction in Audio Language Models},
  pdfauthor={Leonardo Haw-Yang Foo and Hung-yi Lee}
}
\let\ISCSLPthebibliography\thebibliography
\renewcommand{\thebibliography}[1]{\ISCSLPthebibliography{#1}\setlength{\itemsep}{0pt}\setlength{\parsep}{0pt}}

\title{Causal Localization of the Refusal Direction in Audio Language Models}

\name{
  \ifthenelse{\boolean{blind}}{Anonymous to ISCSLP}
  {Leonardo Haw-Yang Foo$^1$, Hung-yi Lee$^{1,2}$}
}

\address{
  \ifthenelse{\boolean{blind}}{Anonymous to ISCSLP}
  {
    $^1$National Taiwan University, Taiwan\\
    $^2$NTU Artificial Intelligence Center of Research Excellence (NTU AI-CoRE), Taiwan
  }
}

\email{
  \ifthenelse{\boolean{blind}}{Anonymous to ISCSLP}
  {leonardofoohy@gmail.com, hungyilee@ntu.edu.tw}
}

\begin{document}

\maketitle

% ===== Abstract =====
\begin{abstract}
A large audio language model (LALM) attaches a speech front end to a text
language model (LM) that is already safety-aligned. When such a model refuses
a harmful spoken request, is the refusal carried by the front end, or
inherited from the text LM? We test this with causal interventions. At each
model's audio-to-LM interface and at tested LM residual layers, we fit a
direction separating harmful from benign prompts, ablate its component, and
measure the resulting change in the model's first-token refusal margin. Four of the five models are
evaluated under held-out category shift. Across five LALMs spanning three
backbone families, with three models passing a baseline safety gate, the largest
tested effects occur in a mid-to-late LM band, while ablations of the tested
interface directions have little effect. On Qwen2.5-Omni, ablating the L16 direction changes the
margin by $-7.10$, versus $-0.013$ at the projector. The audio pathway is
still in use: zeroing the encoder output changes the margin by $-4.7$. In the
same model, the contrast is linearly decodable at an early layer where
single-layer ablation has little effect, and a direction fitted on the LM
backbone alone transfers to the full audio model. Because harmful and benign
prompts also differ in form, we interpret the direction as refusal-linked rather
than harmfulness-specific. These interventions localize dependence of the
refusal margin, not where refusal is computed. Moving the margin also does not
always change what the model writes. Safety audits of these models should use
interventions rather than rely on probes alone, and should examine the
inherited text LM alongside the audio interface.
\end{abstract}

% ISCSLP requires Index Terms after the abstract.

% ===== Body =====
\section{Introduction}
\label{sec:intro}

Large audio language models (LALMs) attach a speech front end to a text LM that
has already been safety-aligned. That architecture leaves a practical ambiguity:
when harmful speech raises refusal logits, is the relevant direction carried by
the audio pathway, or by the inherited LM? The
question matters because speech inputs are a common weak point for LALM
safety~\cite{jalmbench-2025,achilles-2024,acoustic-interference-2026,narrative-audio-2026,ajailbench-2026,emotional-damage-2026}. Throughout, we
measure refusal with one number, $G(x)$: the logit margin between refusal and
compliance tokens at the first output position (\S\ref{sec:method-readout}).
We operationalize the question by asking where
removing a harmful-versus-benign direction makes $G(x)$ fall.

In our experiments, harmful/benign information is decodable from many
activations, but decodability alone does not establish causal
use~\cite{elazar-etal-2021-amnesic,ravichander-2021,belinkov-2022}.
Recent work studies refusal geometry and cross-modal safety dynamics in audio
and multimodal LMs, including alignment and inference-time
interventions~\cite{omni-safety-2026,safety-geometry-collapse-2026,benign-ft-2026,sarsteer-2025,spirit-2025,reshaping-2025}.
To our knowledge, this literature lacks a fixed held-out, site-by-site test
separating where the contrast is decodable from where removing it changes
$G(x)$.

We therefore use direct causal tests. At the audio interface and at LM
residual sites, we fit a rank-1 direction separating harmful from benign inputs
on training categories, then remove it under held-out category shift. The refusal margin depends most
strongly on a mid-to-late LM band, while the tested interface directions have
little effect. Zeroing the encoder output nevertheless shows that the audio
pathway is still in use. On Qwen2.5-Omni, a direction fitted on the LM backbone alone
transfers to the full audio model. These interventions localize causal dependence of the
first-token margin, not where refusal is computed or how generation is
controlled.

\section{Methodology}
\label{sec:method}

Unless otherwise noted, intervention effects are reported as changes in the
first-token margin $G(x)$ (\S\ref{sec:method-readout}); higher $G$ means more
refusal.

\subsection{Datasets and models}
\label{sec:method-setup}
\textbf{Data.} The probe set is 588 matched harmful/benign \emph{audio} pairs
spanning 14 harm categories. Harmful prompts are drawn from three standard text
safety benchmarks: AdvBench~\cite{advbench-2023}, HarmBench~\cite{harmbench-2024},
and JailbreakBench~\cite{jbb-2024}, deduplicated within
category.\footnote{498, 79, and 11 harmful prompts respectively ($588$ total;
$1176$ clips with their matched benign hard negatives), after dropping
within-category near-duplicates at token-Jaccard $\geq 0.8$.}
Each prompt is rendered to speech by a fixed text-to-speech (TTS) voice, so the
model receives the harmful content as audio. Every clip is a \emph{verbatim}
reading of a text prompt, so transcription-mediated transfer remains a standing
caveat; more generally, text priors can obscure whether performance truly
relies on audio~\cite{audio-reliance-2026,audio-text-bias-2025} (\S\ref{sec:lim}).

\textbf{Hard negative.} Each harmful prompt is paired with a \emph{synthesized}
domain-matched hard negative: a benign prompt about the same harm domain, for
example ``how incident responders contain malware'' opposite a malware-authoring
request, generated from a fixed template bank and rendered through the same TTS
pipeline. This pairing matches harm domain and reduces coarse topic
confounding. Its cost is a residual prompt-form difference, templated benign
against natural-imperative harmful, whose footprint we measure directly
(\S\ref{sec:nec}) and discuss (\S\ref{sec:lim}).

\textbf{Held-out.} Two held-out protocols guard against direction-fit
circularity: a \emph{group-blocked} split that holds out four entire unseen harm
categories, and \emph{leave-one-group-out} (LOGO) over 13 of the 14
categories.\footnote{Group-blocked holds out drugs, cyber, bio\_chem, and
fraud\_scam ($N{=}238$ pairs); LOGO aggregates 13 folds ($N{=}584$), omitting
drugs ($N{=}4$, too small for its own fold).} Directions are fit on the
training side of a split and evaluated on the held-out side.

\textbf{Models and sites.} Qwen2.5-Omni-7B~\cite{qwen25omni-2025}
(28 LM decoder layers) is primary. Peers are Audio Flamingo~3~\cite{af3-2025}
(same Qwen2.5-7B backbone, different Whisper variant and projector),
Voxtral-Mini-3B~\cite{voxtral-2025} (30-layer Ministral), and
Ultravox-v0.4.1~\cite{ultravox-2024} and DeSTA2.5~\cite{desta2-2025}
(32-layer Llama-3.1); DeSTA holds the LM \emph{frozen} and feeds it a
Q-Former audio interface plus a spliced ASR transcript. Together the five LALMs
span three LM-backbone families. We probe the relevant front-end interface and
LM residual outputs. Qwen2.5-Omni and AF3 use
$\ell \in \{8, 14, 16, 20\}$; peers use denser early-to-late grids (8, 9, and 9
LM sites for Voxtral, Ultravox, and DeSTA), compared by fractional LM depth.
Interface hooks test selected directions or subspaces at each model's probed
audio-to-LM interface, not the full audio front end, whose overall contribution
is assessed separately.

\subsection{Refusal direction and decodability}
\label{sec:method-dir}
Following Arditi et al.~\cite{arditi-2024}, let
$\mathbf{h}^{(\ell)}_i \in \mathbb{R}^d$ be the residual-stream activation for
prompt~$i$ at layer~$\ell$ and the \emph{readout position}: the last position of
the chat-templated prompt, whose next-token logits define $G(x)$
(\S\ref{sec:method-readout}). With $\mathcal{H}$ and $\mathcal{B}$ the harmful
and benign \emph{training} prompts (audio forwards), the per-layer direction is
the unit-normalized difference of means
\begin{equation}
\mathbf{r}^{(\ell)}
  = \frac{\boldsymbol{\mu}^{(\ell)}_{\mathcal{H}}
        - \boldsymbol{\mu}^{(\ell)}_{\mathcal{B}}}
         {\lVert \boldsymbol{\mu}^{(\ell)}_{\mathcal{H}}
              - \boldsymbol{\mu}^{(\ell)}_{\mathcal{B}}\rVert_2},
\;\;\;
\boldsymbol{\mu}^{(\ell)}_{\mathcal{S}}
  = \frac{1}{|\mathcal{S}|} \sum_{i \in \mathcal{S}} \mathbf{h}^{(\ell)}_i,
\end{equation}
fit independently per layer and per training fold. At the front-end projector,
which has no comparable text readout position, $\mathbf{r}$ is instead the
unit-normalized difference of means over audio-frame positions (mean-pooled per
clip). The rank-$k$ projector check uses this axis plus the top $k{-}1$
principal components of training harmful projector residuals projected onto
$\mathbf{r}^{\perp}$; the resulting basis is orthonormalized and ablated as one
subspace. The audio-derived direction is
used throughout. For the cross-modal transfer analysis (\S\ref{sec:suff}) we
also fit a text counterpart: the same prompts are passed as text tokens to the
LM backbone, bypassing the speech front end, with the direction taken at the
same readout position.

\textbf{Decodability probe.} To separate ``decodable'' from ``causally used,''
we project evaluation prompts onto $\mathbf{r}^{(\ell)}$ and compute
harmful-vs-benign AUROC of that scalar (pair-level 70/30 train/eval split,
in-distribution not group-blocked; eval $N{=}177$ pairs; LM layers only).
AUROC is rank-based, so offsets and scales of raw projections do not matter.
Decodability is correlational by construction; the causal protocol is
\S\ref{sec:method-intervene}--\ref{sec:method-readout}, and \S\ref{sec:probe}
shows the two dissociate.

Because harmfulness and refusal need not share a
direction~\cite{zhao-2025,prompt-safeguard-2024}, and our benign prompts differ in form
(\S\ref{sec:method-setup}), we read $\mathbf{r}^{(\ell)}$ as a refusal-linked
harmful-vs-benign contrast rather than a harmfulness-pure axis
(\S\ref{sec:lim}).

\subsection{Causal intervention}
\label{sec:method-intervene}
Three operations act on a hidden state $\mathbf{x}$ at the hooked module
output:
\begin{equation}
\mathrm{noop}(\mathbf{x}) = \mathbf{x},
\;
\mathrm{ablate}(\mathbf{x}; \mathbf{r}) = \rho\,\mathbf{x}_{\perp},
\;
\mathrm{steer}(\mathbf{x}; \mathbf{r}, k) = \mathbf{x} + k\,\mathbf{r},
\end{equation}
where $\mathbf{x}_{\perp} = \mathbf{x} - (\mathbf{x}^\top \mathbf{r})\mathbf{r}$
removes the rank-1 component along $\mathbf{r}$ and
$\rho = \lVert\mathbf{x}\rVert / \lVert\mathbf{x}_{\perp}\rVert$ rescales to
preserve the activation norm. \textsc{noop} is a plumbing control,
\textsc{ablate} the necessity test, and \textsc{steer} the constructive
companion. Ablation is applied at \emph{every} token position, while
$\mathbf{r}$ is fit at the readout position only: a last-token direction applied
throughout the sequence. The reported add-to-benign test uses one training
mean-shift scale ($\gamma{=}1$), i.e. $k=\lVert
\boldsymbol{\mu}_{\mathcal{H}}-\boldsymbol{\mu}_{\mathcal{B}}\rVert_2$.
A fourth operation, \emph{scalar rescue}, records the prompt's
readout-position pre-ablation coordinate
$a_i=\mathbf{x}_{i,\mathrm{noop}}^\top\mathbf{r}$, ablates, then adds
$a_i\mathbf{r}$ back at all positions; recovery is
$(G_{\mathrm{patch}}-G_{\mathrm{ablate}})/
(G_{\mathrm{noop}}-G_{\mathrm{ablate}})$ per prompt.

To \emph{localize}, we ablate each candidate layer \emph{independently} (one
hooked layer per run), so a site's effect estimates how the held-out margin
depends on removing that layer's fitted direction. This differs from
Arditi et al.~\cite{arditi-2024}, who select a direction at $\ell^\star$ and ablate it at every layer.
Independent ablation asks where the held-out margin
depends on the direction; \S\ref{sec:probe} additionally tests joint
ablation of L08--L15.

\subsection{Readout and localization protocol}
\label{sec:method-readout}
A chat model that refuses almost always begins its answer with an opener
such as ``I'm sorry'' or ``I cannot'', and one that complies begins with an
opener such as ``Sure'' or ``Here''. We therefore read the model's lean toward
refusal off its first output token. Let $\mathcal{R}$ be the token IDs of the
refusal openers and $\mathcal{C}$ those of the compliance openers;\footnote{Full frozen sets (each token with and without a
leading space): refusal $=\{$\textsf{I}, \textsf{I'm}, \textsf{Sorry},
\textsf{Unfortunately}, \textsf{As}, \textsf{Cannot}, \textsf{No},
\textsf{Apolog}$\}$; compliance $=\{$\textsf{Sure}, \textsf{Here},
\textsf{Certainly}, \textsf{Okay}, \textsf{To}, \textsf{Yes}, \textsf{Of}$\}$.}
$G(x)$ is the log-sum-exp margin between the two sets on the first-token logits
$z_t(x)$:
\begin{equation}
G(x) = \log \sum_{t \in \mathcal{R}} e^{z_t(x)} - \log \sum_{t \in \mathcal{C}} e^{z_t(x)},
\end{equation}
so that higher $G$ means a stronger lean toward refusal. The two word lists
were fixed before any intervention was run and are the same for every model;
only the token IDs are re-derived per tokenizer. \textsf{I} is on the refusal
list (and \textsf{I'm} shares its first token) because that is how these
models begin a refusal: in the cached Qwen2.5-Omni generations, all 300
refusals open with ``I'm'', and none of the 180 compliant
answers---including the 60 that ablation induced---begins with \textsf{I}.
Section~\ref{sec:lim} reports how well the sign of $G$ agrees with a
full-text judge. In the results, $\Delta G = G_{\mathrm{ablate}} -
G_{\mathrm{noop}}$ is the change on held-out harmful clips, paired per clip,
and $N$ counts matched pairs in the split; benign, pairwise, and behavioral
results are labeled as such.

\textbf{Specificity nulls.} To check the effect is specific to $\mathbf{r}$ and
not a generic consequence of perturbing the residual stream, we ablate $K$
random unit directions identically at every site\footnote{$K{=}4$ for the
held-out sweeps, $16$ for the peer sweeps, $32$ for the AF3 check, and $2$ for
the rescue, add-induce, and cross-modal analyses.} and compare their
mean effect with that of $\mathbf{r}$. A stricter pre-specified structure-matched null (a direction fit
on benign-vs-benign data, same geometry but no harm contrast) is reported in
\S\ref{sec:lim}.

\textbf{Localization and claim level.} For Qwen2.5-Omni we use
$|\Delta G| \geq 3.55$, half the magnitude of the initial group-blocked L16
effect ($7.10$), as a descriptive threshold for the ``causal band''; dense follow-ups map its extent
(\S\ref{sec:nec}). The primary readout is always first-token $G(x)$, a logit-level
proxy. Behavioral results use greedy 512-token generations scored by the
HarmBench classifier \texttt{cais/HarmBench-Llama-2-13b-cls}~\cite{harmbench-2024},
are labeled separately, and do not follow from logit control
(\S\ref{sec:rep}, \S\ref{sec:lim}). Every unlabeled $\Delta G$ number is
logit-level.

% Start Results in a fresh column: otherwise the first sentence of 3.1 splits
% "pro-/jector" across the page-2/3 boundary and two tables.
\newpage
\section{Results}
\label{sec:results}

\subsection{The margin depends on a mid-LM direction, not on the tested projector direction}
\label{sec:nec}

\begin{figure}[t]
\centering
\includegraphics[width=\columnwidth]{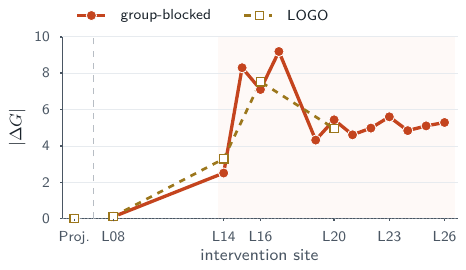}
\caption{\textbf{Ablation localizes the margin effect to an LM band.}
Held-out $|\Delta G|$ at each intervention site, under group-blocked
($N{=}238$) and LOGO ($N{=}584$) category shift; shading marks L14--L26.}
\label{fig:heldout}
\end{figure}

Removing a single direction at a mid-LM layer sharply lowers the refusal
margin; removing the analogous direction at the projector output does not
(Table~\ref{tab:qwen}, Fig.~\ref{fig:heldout}). At layer 16 of
Qwen2.5-Omni, ablation changes the held-out margin by $\Delta G = -7.10$,
against $-0.013$ at the projector, comparable to the random-direction null.
LOGO replicates this contrast; L08 remains small; and the rank-$k$ projector
test of \S\ref{sec:method-dir} reaches at most $2\%$ of the L16 effect at
$k = 16$. Within the projector subspaces we tested, no removal affects the
margin comparably.

\begin{table}[t]
\centering
\caption{\textbf{Qwen2.5-Omni held-out ablation effects.} $\Delta G$ on
harmful clips with paired-bootstrap 95\% CIs, under group-blocked ($N{=}238$)
and leave-one-group-out ($N{=}584$) splits; random-direction nulls satisfy
$|\Delta G| \leq 0.02$ at every site. $\dagger$: dense group-blocked
follow-up.}
\label{tab:qwen}
\scriptsize
\setlength{\tabcolsep}{2.5pt}
\begin{tabular}{@{}lrr@{}}
\toprule
Intervention & Group-blocked & LOGO \\
\midrule
Projector, rank-1 & $-0.013$ [$-0.033$,\,$0.005$] & $-0.005$ [$-0.018$,\,$0.008$] \\
Projector, rank-16 subspace & $-0.14$ [$-0.21$,\,$-0.08$] & --- \\
LM L08 & $-0.10$ [$-0.12$,\,$-0.09$] & $-0.12$ [$-0.14$,\,$-0.11$] \\
LM L14 & $-2.51$ [$-2.64$,\,$-2.39$] & $-3.29$ [$-3.40$,\,$-3.19$] \\
LM L16 & $-7.10$ [$-7.28$,\,$-6.93$] & $-7.53$ [$-7.67$,\,$-7.38$] \\
LM L17$^\dagger$ & $-9.19$ [$-9.38$,\,$-8.99$] & --- \\
LM L20 & $-5.43$ [$-5.56$,\,$-5.27$] & $-4.96$ [$-5.08$,\,$-4.82$] \\
LM L16, natural-benign refit & $-1.79$ [$-1.88$,\,$-1.70$] & --- \\
\bottomrule
\end{tabular}
\end{table}

The effect is not confined to L16 but spread over a band of layers: it rises
from L14, peaks at L15--L18 with its maximum at L17, and is still about half
its maximum at L26 of the model's 28 layers (Table~\ref{tab:qwen},
Fig.~\ref{fig:heldout}). We use L16 below as a representative of this band. Across the other four models the
largest tested effects also occur at LM sites, with model-dependent
mid-to-late profiles, including a broad Voxtral plateau and a DeSTA peak at
L20 (\S\ref{sec:rep}).

\textbf{The audio channel is open: zeroing the encoder output changes $G$ by
$-4.7$ ($N{=}16$ clips).} The projector null is narrower, covering only the
directions and subspaces we tested at that site. Removing a direction inside
the encoder also leaves $G$ unchanged, but only $2$--$3\%$ of the removed
component survives the encoder's final normalization and pooling, so that null
is uninformative (\S\ref{sec:lim}).

\textbf{The size of the L16 effect depends on the benign set but stays far
larger than at L08.} The headline $-7.10$ compares harmful prompts phrased as natural
imperatives with benign prompts generated from templates, so part of it may
reflect prompt form rather than harmful content. Refitting the direction with
natural benign prompts from XSTest~\cite{xstest-2024} reduces the L16 effect
to $-1.79$, below this causal-band threshold of
$|\Delta G| \geq 3.55$ (\S\ref{sec:method-readout}). We therefore treat
$-7.10$ as an effect of the matched design rather than of harmful content
alone. The ordering survives the refit: $|\Delta G|$ remains far larger at
L16 than at L08 ($\Delta G = -0.03$), and random directions again leave $G$
unchanged.

\subsection{Readout-level rescue and text/audio transfer}
\label{sec:suff}

\begin{table}[t]
\centering
\caption{\textbf{Readout-level rescue and transfer at L16 (Qwen2.5-Omni).}
Group-blocked split unless noted. Transfer rows ablate a direction fitted in
one modality during the other modality's forward pass, and give the result as
a fraction of that pass's own effect; the paraphrase row is a fraction of the
verbatim effect on the same $47$ items.}
\label{tab:rescue}
\scriptsize
\setlength{\tabcolsep}{4.5pt}
\begin{tabular}{@{}lrr@{}}
\toprule
Test at L16 & Value & Random control \\
\midrule
Scalar rescue (recovery) & $0.950$ [$0.947$,\,$0.952$] & $0.000$ [$-0.001$,\,$0.001$] \\
Add $\mathbf{r}^{(16)}$ to benign ($\Delta G$) & $+7.24$ & $-0.24$ \\
Text-fit $\to$ audio pass & $0.99$ & --- \\
Audio-fit $\to$ text pass & $0.73$ & --- \\
Paraphrased audio ($N{=}47$) & $0.75$ [$0.65$,\,$0.84$] & --- \\
\bottomrule
\end{tabular}
\end{table}

One scalar coordinate along the L16 direction recovers most of what ablation
removes. Scalar rescue (\S\ref{sec:method-intervene}) records each prompt's
coordinate along $\mathbf{r}^{(16)}$ before ablation and adds it back
afterwards. On the group-blocked split of \S\ref{sec:nec}, ablation carries
the mean margin below zero ($6.85$ to $-0.25$), and the mean per-prompt recovery is
$0.950$, against $0.000$ for a random coordinate (Table~\ref{tab:rescue}). In the opposite direction, adding
$\mathbf{r}^{(16)}$ to benign activations raises the benign margin by
$7.24$.

\textbf{A direction fitted on the LM backbone alone transfers to the full
audio model.} The text-fitted direction, ablated in the audio pass, reproduces
$0.99$ of the effect of the audio-fitted direction at L16; in the reverse
direction the audio-fitted direction reproduces $0.73$ of the text-fitted
effect; the two have cosine similarity $0.76$ (Table~\ref{tab:rescue}). The
transfer is substantial but incomplete: it is asymmetric, and the cosine is
well below one. To check whether shared wording carries the
transfer, we paraphrased the prompts in a separate language-model context,
keeping only those a separate judging pass rated at least $4/5$ for intent
preservation and whose token-level Jaccard overlap with the original was below
$0.3$; the direction was fit on $541$
disjoint pairs and evaluated on $47$ paraphrased clips. The effect is still
$0.75$ of the verbatim-audio effect and remains specific to the fitted
direction (mean random-direction $|\Delta G| \approx 0.045$). Shared wording
alone therefore cannot fully explain the transfer. Because all main clips are verbatim TTS readings, however, the model
may still reuse the text feature through an internal transcript of the speech
(\S\ref{sec:lim}).

\subsection{Decodability appears earlier than single-layer causal dependence}
\label{sec:probe}

A contrast can be easy to decode at a layer where removing it does almost
nothing. The harmful-vs-benign contrast is nearly perfectly decodable at every LM site
we intervene on (AUROC $\geq 0.997$), yet ablating it at L08 changes $G$ by
only $-0.103$, roughly $1/70$ of the L16 effect. Decodability alone
therefore does not localize causal dependence; the templated form of the benign
prompts likely contributes to the ceiling (\S\ref{sec:lim}).

Jointly, however, layers L08--L15 do change the margin. Removing the L08--L15
directions together, each layer with its own training-fit direction on the
same group-blocked split, changes the held-out margin by $-6.01$ (CI
$[-6.24, -5.78]$), or $0.85$ of the single-site L16 effect in the same run
($-7.07$; joint random null $-0.01 \pm 0.07$). One reading is
self-repair~\cite{mcgrath-2023,prakash-2026,selfrepair-2024}: when a single layer's direction is removed,
later layers may restore the margin. But this interval includes L15, whose
own effect is large (Fig.~\ref{fig:heldout}), so this control does not
isolate that reading.
Either way, single-layer ablation localizes margin dependence, not refusal
computation.

\subsection{Cross-backbone replication and the limits of logit control}
\label{sec:rep}

\begin{figure}[t]
\centering
\includegraphics[width=\columnwidth]{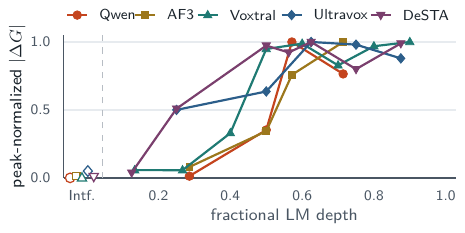}
\caption{\textbf{Layer-wise localization across models.}
For each model, $|\Delta G|$ is normalized by its peak over the tested LM sites
and plotted against fractional LM depth; absolute magnitudes are therefore not
comparable across models. Open markers left of the divider are the tested audio
interfaces; their horizontal positions carry no depth information.}
\label{fig:replication}
\end{figure}

The same LM-band pattern appears in models from all three backbone families
(Table~\ref{tab:xmodel}, Fig.~\ref{fig:replication}). Three of the five
localized models pass our pre-specified safety gate; AF3~\cite{af3-2025} and
Voxtral do not (Table~\ref{tab:xmodel}). A sixth model,
LLaMA-Omni~\cite{fang-etal-2025-llama-omni}, refused almost nothing at baseline
and was excluded before localization. The AF3 direction is also fit in sample
on the $41$ prompts it refuses, so its magnitude is not comparable with the
held-out values.

\begin{table}[t]
\centering
\caption{\textbf{Cross-model summary.} H-refuse is the baseline
harmful-refusal rate; the gate requires it $\geq 0.70$ with answer-benign
$\geq 0.80$, and $\dagger$ marks models that fail it. Harm flip is the
fraction of baseline refusals ablation turns into harmful answers. $\Delta G$
is held-out group-blocked ablation, except AF3, whose in-sample fit makes its
magnitude incomparable; Baseline $G$ and Harm flip come from a separate
behavioral run.}
\label{tab:xmodel}
\scriptsize
\setlength{\tabcolsep}{4pt}
\begin{tabular}{@{}lcccc@{}}
\toprule
Model & H-refuse & Baseline $G$ & LM site ($\Delta G$) & Harm flip \\
\midrule
Qwen2.5-Omni & $1.00$ & $6.85$ & L16 ($-7.10$) & $0.75$ \\
AF3$^{\dagger}$ & $0.68$ & $2.97$ & L20 ($-4.22$) & $0.24$ \\
Voxtral$^{\dagger}$ & $0.60$ & $4.08$ & L18 ($-5.56$) & $0.49$ \\
Ultravox & $0.83$ & $13.12$ & L20 ($-7.69$) & $0.10$ \\
DeSTA & $0.87$ & $8.18$ & L20 ($-6.03$) & $0.00$ \\
\bottomrule
\end{tabular}
\end{table}

The contrast of \S\ref{sec:nec} between LM sites and the tested interface
recurs in the other models: ablation moves the margin by $-0.004$ at the
Voxtral connector, by $+0.40$ at the Ultravox projector against $-7.69$ at
L20, and by $-0.07$ at DeSTA's Q-Former perception output. The shape of the
band differs by family, but in every model the LM sites produce far larger
effects than the interface we tested.

\textbf{Moving the margin does not always change what the model writes.}
Across the four held-out models, band-site ablation turns $0.75$ of Qwen's
baseline refusals into harmful answers ($0.00$ for a random direction),
compared with $0.49$ for Voxtral, $0.10$ for Ultravox, and $0.00$ for DeSTA,
even though the ablation moves the margin by a similar amount in all four
(Table~\ref{tab:xmodel}). Higher baseline margins in Ultravox and DeSTA may
contribute to their smaller harm-flip fractions. However,
Table~\ref{tab:xmodel} reports behavioral baselines and localization-run
$\Delta G$ from separate runs, so it does not establish post-ablation margins
for the behavioral examples.

\section{Limitations and Discussion}
\label{sec:lim}

We do not claim that refusal is computed in the LM. We show where the
first-token margin $G(x)$ falls when a single direction is removed; where
refusal is computed, and whether moving the margin controls what the model
generates, are separate questions.

$G(x)$ reads only the first token, and a first token that leans toward
compliance does not guarantee a harmful answer: the fraction of refusals that
ablation turns into harmful answers ranges from $0.75$ to $0.00$ across models
(Table~\ref{tab:xmodel}), all scored by one HarmBench classifier. On
Qwen2.5-Omni, Claude Opus 4.8 labeled the $120$ unablated outputs and the
HarmBench classifier the $60$ ablated harmful outputs. The sign of $G$ agreed
with these labels on $0.70$ of the ablated outputs, whose margins lie near
zero; on $0.81$ of all $180$ items (Cohen's $\kappa = 0.62$); and on $0.97$ of
items with $|G| > 3$. This mixed-evaluator check has not been repeated on the
other models.

The projector null applies to the tested directions and subspaces, not to the
audio pathway.
Zeroing the whole audio encoder moves $G$ by $-4.7$. No single direction fit
on mean-pooled projector frames, and no rank-$k$ subspace up to $k{=}16$,
changes $G$ by more than $2\%$ of the L16 effect when removed. Distributed,
position-specific, or nonlinear mechanisms in the audio pathway are not ruled
out~\cite{wollschlager-2025}.

A pre-specified structure-matched specificity check was
ambiguous under its decision rule: across eight benign-versus-benign splits,
mean $|\Delta G|$ was $0.47$ (maximum $1.89$), above the isotropic
random-direction band but small relative to the fitted-direction effect. For the VITS and
SpeechT5 voices, XSTest-safe prompts yield $\Delta G = -2.86$ and $-2.93$,
versus $-6.75$ and $-7.58$ on the corresponding unsafe contrasts. Together these
controls argue against a harmfulness-pure reading: we interpret the direction
as refusal-linked rather than harmfulness-specific.

All speech is synthetic: on Qwen2.5-Omni the contrast between L16 and the
projector reproduces across three further TTS voices, but human-recorded
speech remains untested. The main next steps are to test human-recorded
speech and audio mechanisms not mediated by a transcript.

\section{Conclusion}
\label{sec:conc}

We asked whether refusal in audio language models is carried by the speech
front end or by the inherited language-model backbone. Under the tested rank-1
intervention, the first-token refusal margin $G(x)$ depends most strongly on a
mid-to-late LM residual band, while the tested projector/connector directions
stay near zero. This localizes dependence of the refusal readout, not
where refusal is computed; it does not make the audio pathway irrelevant. For safety auditing, decodability alone does not establish
causal use; inherited text-LM backbones should be audited alongside audio
interfaces.

\par
\noindent\textbf{Ethics statement.}
We report aggregate white-box safety measurements only and release no fitted
directions, prompts, or generations.

% ===== Acknowledgements (anonymous version: omitted under blind) =====
\ifthenelse{\boolean{blind}}{}{%
  \section{Acknowledgments}
  This work was supported by the Ministry of Education (MOE) of Taiwan under
  the project Taiwan Centers of Excellence in Artificial Intelligence, through
  the NTU Artificial Intelligence Center of Research Excellence (NTU AI-CoRE).
  We thank the National Center for High-performance Computing (NCHC) of the
  National Applied Research Laboratories (NARLabs) in Taiwan for providing
  computational and storage resources.
}

% ===== References =====
\bibliographystyle{IEEEtran}
\bibliography{references}

\begin{thebibliography}{10}
\providecommand{\url}[1]{#1}
\csname url@samestyle\endcsname
\providecommand{\newblock}{\relax}
\providecommand{\bibinfo}[2]{#2}
\providecommand{\BIBentrySTDinterwordspacing}{\spaceskip=0pt\relax}
\providecommand{\BIBentryALTinterwordstretchfactor}{4}
\providecommand{\BIBentryALTinterwordspacing}{\spaceskip=\fontdimen2\font plus
\BIBentryALTinterwordstretchfactor\fontdimen3\font minus
  \fontdimen4\font\relax}
\providecommand{\BIBforeignlanguage}[2]{{%
\expandafter\ifx\csname l@#1\endcsname\relax
\typeout{** WARNING: IEEEtran.bst: No hyphenation pattern has been}%
\typeout{** loaded for the language `#1'. Using the pattern for}%
\typeout{** the default language instead.}%
\else
\language=\csname l@#1\endcsname
\fi
#2}}
\providecommand{\BIBdecl}{\relax}
\BIBdecl

\bibitem{jalmbench-2025}
Z.~Peng \emph{et~al.}, ``{JALMBench}: Benchmarking jailbreak vulnerabilities in
  audio language models,'' arXiv preprint arXiv:2505.17568, 2025.

\bibitem{achilles-2024}
H.~Yang, L.~Qu, E.~Shareghi, and G.~Haffari, ``Audio is the achilles' heel: Red
  teaming audio large multimodal models,'' arXiv preprint arXiv:2410.23861,
  2024.

\bibitem{acoustic-interference-2026}
Y.~Wang, Y.~Huang, Z.~Liang \emph{et~al.}, ``Acoustic interference: A new
  paradigm weaponizing acoustic latent semantic for universal jailbreak against
  large audio language models,'' arXiv preprint arXiv:2605.18168, 2026.

\bibitem{narrative-audio-2026}
Y.~Yu, H.~Jin, Y.~Yu, J.~Zhuang, and H.~Wang, ``Now you hear me: Audio
  narrative attacks against large audio--language models,'' in
  \emph{Proceedings of the 19th Conference of the European Chapter of the
  Association for Computational Linguistics (Volume 1: Long Papers)}.\hskip 1em
  plus 0.5em minus 0.4em\relax Association for Computational Linguistics, Mar.
  2026, pp. 5925--5939.

\bibitem{ajailbench-2026}
Z.~Song, Q.~Jiang, M.~Cui \emph{et~al.}, ``Audio jailbreak: An open
  comprehensive benchmark for jailbreaking large audio-language models,'' in
  \emph{Proceedings of the 64th Annual Meeting of the Association for
  Computational Linguistics (Volume 1: Long Papers)}.\hskip 1em plus 0.5em
  minus 0.4em\relax Association for Computational Linguistics, 2026, pp.
  27\,294--27\,308.

\bibitem{emotional-damage-2026}
B.-H. Feng, C.-F. Liu, Y.-H.~L. Liang \emph{et~al.}, ``Emotional damage:
  Investigating safety vulnerabilities of large audio-language models under
  speaker emotional variations,'' in \emph{ICASSP 2026 -- 2026 IEEE
  International Conference on Acoustics, Speech and Signal Processing
  (ICASSP)}.\hskip 1em plus 0.5em minus 0.4em\relax IEEE, 2026, pp.
  17\,792--17\,796.

\bibitem{elazar-etal-2021-amnesic}
Y.~Elazar, S.~Ravfogel, A.~Jacovi, and Y.~Goldberg, ``Amnesic probing:
  Behavioral explanation with amnesic counterfactuals,'' \emph{Transactions of
  the Association for Computational Linguistics}, vol.~9, pp. 160--175, 2021.

\bibitem{ravichander-2021}
A.~Ravichander, Y.~Belinkov, and E.~Hovy, ``Probing the probing paradigm: Does
  probing accuracy entail task relevance?'' in \emph{Proceedings of the 16th
  Conference of the European Chapter of the Association for Computational
  Linguistics: Main Volume}.\hskip 1em plus 0.5em minus 0.4em\relax Association
  for Computational Linguistics, 2021, pp. 3363--3377.

\bibitem{belinkov-2022}
Y.~Belinkov, ``Probing classifiers: Promises, shortcomings, and advances,''
  \emph{Computational Linguistics}, vol.~48, no.~1, pp. 207--219, 2022.

\bibitem{omni-safety-2026}
K.~Wang, Z.~Li, Z.~Zhou \emph{et~al.}, ``{Omni-Safety} under cross-modality
  conflict: Vulnerabilities, dynamics mechanisms and efficient alignment,''
  arXiv preprint arXiv:2602.10161, 2026.

\bibitem{safety-geometry-collapse-2026}
J.~Guo, X.~Guo, J.~Chen \emph{et~al.}, ``Safety geometry collapse in multimodal
  {LLMs} and adaptive drift correction,'' arXiv preprint arXiv:2605.18104,
  2026.

\bibitem{benign-ft-2026}
J.~Roh and A.~Houmansadr, ``Benign fine-tuning breaks safety alignment in audio
  {LLMs},'' arXiv preprint arXiv:2604.16659, 2026.

\bibitem{sarsteer-2025}
W.~Lin, J.~Li, H.~Xiong \emph{et~al.}, ``{SARSteer}: Safeguarding large
  audio-language models via safe-ablated refusal steering,'' arXiv preprint
  arXiv:2510.17633, 2025.

\bibitem{spirit-2025}
A.~Djanibekov, N.~Mukhituly, K.~Inui \emph{et~al.}, ``{SPIRIT}: Patching speech
  language models against jailbreak attacks,'' arXiv preprint arXiv:2505.13541,
  2025.

\bibitem{reshaping-2025}
H.~Yang, L.~Qu, E.~Shareghi, and G.~Haffari, ``Reshaping representation space
  to balance the safety and over-rejection in large audio language models,'' in
  \emph{Proceedings of the 2025 Conference on Empirical Methods in Natural
  Language Processing}.\hskip 1em plus 0.5em minus 0.4em\relax Suzhou, China:
  Association for Computational Linguistics, 2025, pp. 10\,067--10\,079.

\bibitem{advbench-2023}
A.~Zou, Z.~Wang, N.~Carlini \emph{et~al.}, ``Universal and transferable
  adversarial attacks on aligned language models,'' arXiv preprint
  arXiv:2307.15043, 2023.

\bibitem{harmbench-2024}
M.~Mazeika, L.~Phan, X.~Yin \emph{et~al.}, ``{HarmBench}: A standardized
  evaluation framework for automated red teaming and robust refusal,'' arXiv
  preprint arXiv:2402.04249, 2024.

\bibitem{jbb-2024}
P.~Chao, E.~Debenedetti, A.~Robey \emph{et~al.}, ``{JailbreakBench}: An open
  robustness benchmark for jailbreaking large language models,'' arXiv preprint
  arXiv:2404.01318, 2024.

\bibitem{audio-reliance-2026}
L.~H.-Y. Foo, C.-K. Yang, C.-A. Li, K.-H. Lu, and H.-y. Lee, ``All that
  glitters is not audio: Rethinking text priors and audio reliance in
  audio-language evaluation,'' arXiv preprint arXiv:2604.24401, 2026.

\bibitem{audio-text-bias-2025}
C.~Wang, G.~Deng, X.~Yang, H.~Qiu, and T.~Zhang, ``When audio and text
  disagree: Revealing text bias in large audio-language models,'' in
  \emph{Proceedings of the 2025 Conference on Empirical Methods in Natural
  Language Processing}.\hskip 1em plus 0.5em minus 0.4em\relax Suzhou, China:
  Association for Computational Linguistics, 2025, pp. 4878--4888.

\bibitem{qwen25omni-2025}
J.~Xu, Z.~Guo, J.~He \emph{et~al.}, ``{Qwen2.5-Omni} technical report,'' arXiv
  preprint arXiv:2503.20215, 2025.

\bibitem{af3-2025}
A.~Goel, S.~Ghosh, J.~Kim \emph{et~al.}, ``{Audio Flamingo 3}: Advancing audio
  intelligence with fully open large audio language models,'' arXiv preprint
  arXiv:2507.08128, 2025.

\bibitem{voxtral-2025}
A.~H. Liu, A.~Ehrenberg, A.~Lo \emph{et~al.}, ``{Voxtral},'' arXiv preprint
  arXiv:2507.13264, 2025.

\bibitem{ultravox-2024}
{Fixie AI}, ``{Ultravox},'' \url{https://github.com/fixie-ai/ultravox}, 2024,
  speech-language model, v0.4.1 (Llama-3.1-8B backbone).

\bibitem{desta2-2025}
K.-H. Lu, Z.~Chen \emph{et~al.}, ``{DeSTA2.5-Audio}: Toward general-purpose
  large audio language model with self-generated cross-modal alignment,'' arXiv
  preprint arXiv:2507.02768, 2025.

\bibitem{arditi-2024}
A.~Arditi, O.~Obeso, A.~Syed, D.~Paleka, N.~Panickssery, W.~Gurnee, and
  N.~Nanda, ``Refusal in language models is mediated by a single direction,''
  2024, also appeared at NeurIPS 2024.

\bibitem{zhao-2025}
J.~Zhao, J.~Huang, Z.~Wu \emph{et~al.}, ``{LLMs} encode harmfulness and refusal
  separately,'' arXiv preprint arXiv:2507.11878, 2025.

\bibitem{prompt-safeguard-2024}
C.~Zheng, F.~Yin, H.~Zhou \emph{et~al.}, ``On prompt-driven safeguarding for
  large language models,'' in \emph{Proceedings of the 41st International
  Conference on Machine Learning}, ser. Proceedings of Machine Learning
  Research, vol. 235.\hskip 1em plus 0.5em minus 0.4em\relax PMLR, 2024, pp.
  61\,593--61\,613.

\bibitem{xstest-2024}
\BIBentryALTinterwordspacing
P.~R{\"o}ttger, H.~Kirk, B.~Vidgen, G.~Attanasio, F.~Bianchi, and D.~Hovy,
  ``{XST}est: A test suite for identifying exaggerated safety behaviours in
  large language models,'' in \emph{Proceedings of the 2024 Conference of the
  North American Chapter of the Association for Computational Linguistics:
  Human Language Technologies (Volume 1: Long Papers)}.\hskip 1em plus 0.5em
  minus 0.4em\relax Mexico City, Mexico: Association for Computational
  Linguistics, Jun. 2024, pp. 5377--5400. [Online]. Available:
  \url{https://aclanthology.org/2024.naacl-long.301/}
\BIBentrySTDinterwordspacing

\bibitem{mcgrath-2023}
T.~McGrath, M.~Rahtz, J.~Kram{\'a}r, V.~Mikulik, and S.~Legg, ``The hydra
  effect: Emergent self-repair in language model computations,'' arXiv preprint
  arXiv:2307.15771, 2023.

\bibitem{prakash-2026}
N.~Prakash, W.~J. Yeo, A.~Abdullah, R.~Satapathy, E.~Cambria, and R.~K.-W. Lee,
  ``Beyond {I}'m sorry, {I} can't: Dissecting large-language-model refusal,''
  in \emph{Proceedings of the AAAI Conference on Artificial Intelligence},
  vol.~40, no.~44, 2026, pp. 37\,830--37\,838.

\bibitem{selfrepair-2024}
C.~Rushing and N.~Nanda, ``Explorations of self-repair in language models,'' in
  \emph{Proceedings of the 41st International Conference on Machine Learning},
  ser. Proceedings of Machine Learning Research, vol. 235.\hskip 1em plus 0.5em
  minus 0.4em\relax PMLR, 2024, pp. 42\,836--42\,855.

\bibitem{fang-etal-2025-llama-omni}
\BIBentryALTinterwordspacing
Q.~Fang, S.~Guo, Y.~Zhou, Z.~Ma, S.~Zhang, and Y.~Feng, ``{LLaMA-Omni}:
  Seamless speech interaction with large language models,'' in \emph{The
  Thirteenth International Conference on Learning Representations}, 2025.
  [Online]. Available: \url{https://openreview.net/forum?id=PYmrUQmMEw}
\BIBentrySTDinterwordspacing

\bibitem{wollschlager-2025}
T.~Wollschl{\"a}ger, J.~Elstner, S.~Geisler, V.~Cohen-Addad, S.~G{\"u}nnemann,
  and J.~Gasteiger, ``The geometry of refusal in large language models: Concept
  cones and representational independence,'' in \emph{Proceedings of the 42nd
  International Conference on Machine Learning}, ser. Proceedings of Machine
  Learning Research, vol. 267.\hskip 1em plus 0.5em minus 0.4em\relax PMLR,
  2025, pp. 66\,945--66\,970.

\end{thebibliography}

\end{document}